# *The general theory of species abundance distributions*

Michael G Bowler[1] and Colleen K Kelly[2]

[1]Department of Physics, University of Oxford, Keble Road, Oxford OX1 3RH, UK; Email: *michael.bowler@physics.ox.ac.uk*; [2]Department of Zoology, University of Oxford, South Parks Road, Oxford OX1 3PS, UK; Email: *colleen.kelly@zoo.ox.ac.uk*

**Abstract.** A central issue in ecology today is that of the factors determining the relative abundance of species within a natural community. The proper application of the principles of statistical physics to the problem of species abundance distributions (SADs) has enabled us to identify the fundamental ecological principles responsible for the near universal features observed. These principles are (i) a limit on the number of individuals in an ecological guild and (ii) *per capita* birth and death rates. We thus unify the neutral theory of Hubbell [1], the master equation approach of Volkov et al [2, 3] and a recent attempt at deploying statistical mechanics to deal with this problem, the idiosyncratic [extreme niche] theory of Pueyo et al [4]; we have identified the true origin of the prior in [4]. We also find in our results clear indications that niches must be very flexible and that temporal fluctuations on all sorts of scales are of considerable importance in community structure.

## 1. Introduction

The species constituting ecological communities are represented by different numbers of individuals; generally there are many species with relatively few individuals and relatively few species with many individuals. The distribution of species over the number of individuals per species, species abundance distributions or SADs, are recognisably drawn from a single family of distributions, ranging from the log series distribution (for example, for marine diatoms [5], through the log normal distributions more or less characteristic of tropical forests [1, 2, 6], to the highly skewed and unveiled log normal characterising British birds [7]. The existence of such a family of curves (see for example Fig 1.1 of [1]) is impressive and it seems obvious that very general ecological principles must govern species abundance. It is our contention that the proper application of the ideas of statistical mechanics, familiar in the physical sciences, has allowed us to identify these principles. There are two – (i) that the most important constraint is that applied to the total number of individuals in the guild (the sum of the populations of the individual species in the guild) and (ii), that individuals give birth and die. More specifically, at least to a good approximation, the dynamics are controlled by *per capita* birth rates and death rates. Constraints on the populations of individual species or correlations between species are not of dominant importance in determining the species abundance distribution; the successes of neutral theory emerge demonstrably as an appropriate average over species dynamics and the closer reality is to the idiosyncratic end of the spectrum of models [4], the more likely it is that temporal fluctuations at all scales are important in averaging out differences between individual species.

The log series distribution, at one extreme of the SAD spectrum and a good place to start, is

$$\psi(n) \propto \exp(-\beta n)/n \qquad (1)$$

In the proper application of statistical mechanics, the exponential emerges from the constraint (i) on the total number of individuals and the factor $1/n$ is the direct result of the very biological processes of individuals giving birth and dying. These are our essential results and they emerge as a development of the most interesting (but not entirely convincing) paper of Pueyo et at [4]. We drew on the master equation approach of Volkov et al [2, 3], in light of our knowledge of statistical mechanics [8] and certain of our earlier results by no means purely theoretical [6, 9, 10].

## 2. Application of statistical mechanics to population problems

We would first emphasise that statistical mechanics is the calculation of emergent properties of very complex dynamical systems subject to fairly straightforward underlying principles. It is expressed in terms of probabilities because it inevitably involves some sort of averaging and as a result some of the techniques and too much of the verbal language is also encountered in information theory. This can lead to much confusion: in particular the (Shannon) information theory entropy is a different thing from the thermodynamic entropy encountered in statistical mechanics in the physical sciences (emphasised by Jaynes [11] pg. 351). We prefer to develop our argument in terms of biological entities as real as atoms and processes as real as the scattering of atoms, rather than deploying information theory.

One can approach an equilibrium configuration of a complex system by recognising that since such a system is dynamical it is continually exploring all possible configurations. Thus in some sort of equilibrium the most probable configuration is most likely to be encountered (overwhelmingly so if there are $10^{23}$ constituents). Another way is to set up a dynamical equation governing the evolution of probabilities (or numbers in some class) as a function of time; a dynamical equation which converges to an equilibrium (probabilistic) solution. Both approaches are enlightening (a physicist might compare lectures 2 [the former] and 3 [the latter] in [8]). We start by applying the statistical mechanics of the most probable configuration to the problem of species abundance.

### 2a. Most probable configuration

This approach to species abundance goes back at least to MacArthur [12, 13]. It is most usefully presented here as a combinatorial argument.

Suppose we have $S$ objects (they are going to be species) assigned to classes such that the class labelled $n$ contains a number $s_n$ objects. In the context of species abundance, there are $s_n$ species with population $n$ individuals in each. The number of ways of arranging $S$ objects over the different classes so as to achieve a configuration $\{s_n\}$, characterised by the numbers in each class $s_1, s_2, ... s_n, ...$, is given by

$$W = \frac{S!}{\Pi s_n!} \qquad (2)$$

where $\Pi$ denotes the continued product. This combinatoric argument is of course completely independent of the nature of the objects and the classes into which they are to be placed. The quantity $W$ is proportional to the probability of finding this particular configuration $\{s_n\}$ provided that each and every arrangement has, without further conditions being imposed, equal weight. At this level every object has the same probability of being found in every class. That this is so constitutes a biological assumption which may or may not be suitable for calculating a most probable configuration for the distribution of species over populations. If this is not true, an additional weight factor expressing unequal probabilities can be introduced into (2); a factor containing *priors*. Start by keeping the assumption that the weight $W$ is indeed proportional to the number of micro states realising $\{s_n\}$ – that the prior distribution is uniform in $s_n$ - and hence may be maximised to yield the most probable configuration, provided that the maximisation is done subject to the real physical or biological constraints relevant to the problem. The first constraint is a normalisation condition; we are dealing with a fixed number of species $S$ and wish to determine how they will distribute themselves when subject to the second condition, that the total number of individuals $N$ is fixed (for example, there are 273 species of tree in the forest and 12,241 mature individuals). Then the first condition is the trivial one we started out with:

$$\sum s_n = S \qquad (3a)$$

and the second is

$$\sum n s_n = N \qquad (3b)$$

In practice the logarithm of (2) is maximised subject to the constraints (3) and the standard technique is to use Lagrange multipliers (see [14]). The solution is very simple

$$s_n = s_0 \exp(-\beta n) \qquad (4)$$

In (4), the constants $s_0$ and $\beta$ are determined by the conditions in (3).

As a candidate distribution for species abundance, this is a dismal failure; it will be familiar to the general ecological reader as MacArthur's broken stick [12] (where "broken" refers to a pictorial representation of the model, not to its SAD failure). It does not have the form (1), nor anything close to it, but note that the exponential arises from the condition imposed on the total number of individuals $N$. Recent attempts at applying statistical mechanics to the problem of species abundance have sought to justify the introduction of the additional vital factor of $1/n$. These attempts are a biologically grounded model on the one hand [15]; on the other, postulates that maximally uninformative priors (in the sense of information theory) should be introduced [4, 16]. We agree that Pueyo et al [4] have correctly identified the prior distribution (and with much else in that interesting paper) but we disagree with their appeal to some maximally uninformative prior: the factor of $1/n$ can be straightforwardly traced to the role of birth and death of individuals in ecological systems (our second principle; (ii) above). There is then a sense in which this prior is highly informative.

If the combinatorial argument constructed above is to be in any way applicable to species abundance, the flaw in the reasoning that leads to (4) might be that there are other constraints which are important (the model of Harte et al [15] is an ingenious attempt), but more likely the flaw is the assumption that any class labelled by $n$ is, in the absence of constraints, an equally probable destiny for species. This emerges as the correct approach from our work below, and we give two examples of how dynamics can make a difference. First, suppose that there are $g_n$ distinct classes with population $n$ and all distinct classes are equally probable. Then the number of species with population $n$ is going to be given by multiplying (4) by the factor $g_n$ and (1) would be recovered if $g_n = 1/n$. This is a concrete example of how the assumptions underlying the application of statistical mechanics could be modified, but we do not seriously advance this model as a solution to the problem of species abundance. In a different context it is not ridiculous at all – in quantum mechanics it frequently occurs that there are several energy levels sharing the same energy but otherwise distinct and equally open to occupation by atoms. The phenomenon is called degeneracy and must be taken into account in properly turning statistical mechanics into the kinetic theory of gases [Appendix A].

Our second example applies to the statistical mechanics of species abundance distributions. Return to (2) and now suppose not that there are $g_n$ separate classes each characterised by *n*, but rather that each (single) class has, in the absence of the constraints (3), not equal probability of containing a species but rather a probability proportional to $g_n$. The origin of such a factor is not to be sought in counting multiple states, but rather in the underlying dynamics of species sorting out abundances. It is this factor which is due to the dynamics of birth and death of individuals; we shall turn to its origin in those dynamics shortly. The effect is of course the same as that of the multiple states hypothesis; namely with the inclusion of this factor the solution (4) is multiplied by $g_n$. In either case, it is appropriate to maximise a modified weight factor

$$W^g = \frac{S!}{\Pi s_n!} \Pi g_n^{s_n} \qquad (5)$$

because the probability of there being $s_n$ (species) in the class characterised by (population) *n* is increased by a relative factor of $g_n^{s_n}$. Compare (5) with eq.(8) of Pueyo et al [4]: their prior factor $P_\pi(n)$ is matched by our $g_n$. We have suggested above two possible concrete or mechanistic origins for such a factor, whereas Pueyo et al [4] appeal to a supposed principle that the entropy should be maximised relative to some minimally informative prior distribution extracted via an invariance. This has always seemed a trifle mysterious; our argument below is rooted in the dynamics. We note that the $1/n$ prior is equivalent to a prior distribution uniform in the number of individuals summed over all species, $ns_n$.

The processes of birth and death of individuals provide the mechanism which automatically ensures that the probability of a species being in the class of population *n* contains a factor $g_n = 1/n$. One might say that these dynamics give us prior information which must be included (by discarding (2) in favour of (5)) before maximising subject to the constraints (3). We find that the prior ($1/n$) employed by Pueyo et al [4] is essentially correct, but their reasoning is wrong.

The biological machinery whereby birth and death dictate (5), with $g_n = 1/n$, as opposed to (2), is very simple. A species in the class characterised by population *n* transits out of this class either by giving birth to a new individual $(n \rightarrow n+1)$ or by an individual dying $(n \rightarrow n-1)$. The rate at which the species vacates this class is the sum of the birth rate and the death rate and with reasonably well defined *per capita* rates the exit rate is proportional to the population *n*. Thus the prior probability for a species to be in a class characterised by *n* contains a factor $g_n = 1/n$. This argument conveys the essence in a verbal form; rigour is to be found in the mathematical argument developed below and enlarged upon in appendix B.

## 2b. A dynamical approach

The largely verbal argument given in the previous section may be persuasive but is not complete. Our result is established more rigorously through equations which embody the underlying dynamics.

The number of species in the class with population *n* is depleted if an individual in any species in that class dies. It is also depleted if any individual gives birth to another individual. There are $ns_n$ individuals in the class and so, given *per capita* birth and death rates, the depletion rate for the number of species in class *n* is given by

$$-(\overline{b_n} + \overline{d_n})ns_n$$

Similarly, if an individual in the class with population $n-1$ gives birth, the number of species in class *n* increases by 1; if an individual in a species in class $n+1$ dies, then again the number of species in the class with population *n* increases. Thus the rate at which the number of species in class *n* is increased is given by

$$+\overline{b_{n-1}}(n-1)s_{n-1} + \overline{d_{n+1}}(n+1)s_{n+1}$$

Putting the two together yields a differential equation coupling the numbers of species with $n-1$, *n*, $n+1$ individuals. The *per capita* birth and death rates $\overline{b}$ and $\overline{d}$ are clearly averages over species and in terms of these averages we have

$$\frac{ds_n}{dt} = \overline{b_{n-1}}(n-1)s_{n-1} - \overline{d_n}ns_n - \overline{b_n}ns_n + \overline{d_{n+1}}(n+1)s_{n+1} \qquad (6)$$

which is a version of the master equation encountered in Volkov et al [2, 3]. If we take the extremely simple case where all the average *per capita* birth and death rates are the same, then the solution for the steady state (where the left hand side of (6) vanishes for all *n*) is, apart from a normalising constant,

$$s_n = \frac{1}{n}\left(\frac{\overline{b}}{\overline{d}}\right)^n \qquad (7)$$

which is of the form (1). This is easily verified by direct substitution, and the origin of the factor of $1/n$ in the *per capita* birth and death rates is also easily verified by attributing *per capita* rates to species rather than to individuals; replacing the factors of $n-1$, *n*, $n+1$ in (6) by unity. A distribution equivalent to (4) then results (no $1/n$).

There is a separate ecological problem for which this different assumption is entirely appropriate and exponential distributions rather than log series are encountered. In the distribution of alien species over sites world wide, the number of species found at *n* sites falls exponentially with *n* and similarly for the number of pairs of sites sharing *n* species [9]. In that problem we envisage suitable niches opening and closing as time and climate evolve, but the chance of a niche opening or closing at any particular site will be independent of the number of alien species already present; an alien species does not give rise to a new alien species.

At this point we have established our contention that the fundamental ecological properties responsible for the form of species abundance distributions are first that the total number of individuals is fixed and secondly that birth and death rates are *per capita*. It remains to consider the effects of variants about the most simple assumptions and also the implications of these two principles – which are hardly contentious.

**3. Neutrality, symmetry, idiosyncracy?**

The log series distribution represented by (1) and also by (7) is the simplest realisation of a species abundance distribution governed primarily by the two principles enunciated earlier. The log series is at one extreme end of the distributions observed but it is a very good place to start.. Departures in the direction of log normal distributions can easily be made [3]. Pueyo et al [4] introduce additional constraints in terms of ln *n* and its square, whereas Volkov et al [2] introduce departures from pure *per capita* birth and death rates, making the $\bar{b}$ and $\bar{d}$ in (6) mildly dependent on the population *n,* so as to allow for density dependent effects of the kind that exert a stabilising effect on species coexistence. An additional constraint, on resource usage, was introduced by Dewar & Porte [16] and by Harte et al [15]. We are not here concerned with these details and refer the reader directly to those works. Here we are primarily concerned with the circumstances under which maximisation of the weight $W^g$ in (5) is appropriate and establishing a connection with conditions which make (6) a valid approximation. The classes of ecological theory we consider are neutrality [1], symmetry [2, 17] and idiosyncracy [4].

The case of a neutral theory (all individuals in a guild, regardless of species, are identical) is a special case of symmetric theories (all individuals are equal but some are more equal than others but all species are identical) whereas an idiosyncratic theory treats every species as essentially different from every other. It is that latter case that requires the most attention. As far as equation (5) is concerned, if all species are identical then all species have the same probability of joining the class defined by *n* and (5) is valid. In terms of the dynamical equation (6), in a neutral theory *b* and *d* are the same for all species and independent of the population *n*; in a symmetric theory the birth and death rates *b* and *d* would strictly be the same for all species, but (6) would clearly correspond to a form of species symmetry if merely the ratios *b/d* were the same for all species. The interesting case is that where species are taken as being very different; the idiosyncratic model of Pueyo et al [4]. We give below our summary of the essential idea.

*3a. The nature of idiosyncracy*

Pueyo et al [4] consider the possibility that a guild is composed of species each of which is governed by a distinct ecological model and that there are a very large number of such models. An arbitrary species might be described by any such model and the probability that a species labelled by the numeral 1 has population $n_1$ is an appropriate average over all possible models, $P_\pi(n_1)$. The same distribution applies to a second arbitrarily chosen species and so given such a universal function it follows that (5) will be valid in this extreme idiosyncratic limit, with $g_n$ playing the part of $P_\pi(n)$. They argue that the form of $P_\pi(n)$ should be such that " the bits of information that are different in different models cancel out. If we include all conceivable models, the resulting distribution $P_\pi$ will be completely void of information (any bit of information surviving … would mean some ecological feature that systematically appears in many different species and that needs explanation)". Our dynamical reasoning, which led us to select a factor $1/n$ for $P_\pi$, exactly matches this verbal reasoning of Pueyo et al; all conceivable ecological models have, because of the rules for birth and death, a $1/n$ bias against a species being populous and so this bit of information should not cancel out and it has a clear explanation. In the language of information theory, we have a prior $P_\pi(n) = 1/n$. This contains information and yet contains no information outside of this common necessity. In our derivation involving dynamics it appears as a weight factor with a clear biological origin and is no more mysterious than a degeneracy factor due to many different quantum states having the same energy. We do not otherwise dispute the most important paper of Pueyo et al [4] and we agree that their choice of $P_\pi$ is the right one - for species abundance distributions.

### *3b. Neutrality, symmetry, idiosyncrasy and the dynamical approach*

Equation (6) is an adaptation of equation (1) of Volkov et al [3]. Their equations were originally written as equations for the evolution with time of the probability that a species $k$ has population $n$. Our (6) is an average over the equations for individual species and clearly this will be valid for models which are symmetric (including neutral models) because the label $k$ is redundant. It is important to note that, in general, birth and death rates may be species specific, vary with the population $n$ and indubitably are affected by the populations of other species and interactions between species (an example in terms of the statistical mechanics of gases is given in Appendix A). Thus in general the different $b_n^k(t)$ can be very different and a function of time; similarly for the *per capita* death rates $d_n^k(t)$. Idiosyncracy must be embodied in these parameters when expressing the problem in dynamical terms and as equilibrium (a dynamical equilibrium) is approached, the averaging process which gives rise to (6) will include not only averaging over species but also over a period of time (long in comparison with the timescale for birth and death of individuals but short in comparison with the relaxation time for species abundance near equilibrium). This is discussed in more detail in Appendix B; here we note that the expression of the idiosyncracy in terms of the dynamical equation (6) must be that departures in individual species from the averaged picture which is (6) must become unimportant on averaging over all species involved and over the appropriate time scale. We note here the potentially very important effect on community structure that could be played by temporal fluctuations on all sorts of time scales. Temporal effects have recently been attracting some attention, primarily in the context of stable coexistence of a few species (even though embedded in a complex community) [18-23], but not really for the structure of complex communities containing maybe several hundred species (but see [24]). It may be hard to disentangle the effects of temporal structure, but as a promoter of idiosyncracy it looks very promising.

### 4. Conclusions

A simple combination of ideas on community structure which have emerged over the last decade, illuminated by our knowledge of statistical mechanics, has led to an understanding of the essential features of species abundance distributions in terms of two general ecological principles: (i) The number of individuals in a community of species sufficiently similar as to constitute a guild is fixed (more or less) and (ii), that the dynamics of birth and death operate on a *per capita* basis; individuals die and individuals give birth to new individuals.

This is in no sense to assert that there are no other ecological interactions, merely that they are of minor importance in determining the species abundance distribution. The results are most easily appreciated in the context of a neutral or symmetric theory of community structure but can equally well emerge as a result of averaging over communities where species may individually depart considerably from the mean behaviour [4]. It has long been suspected that the successes of neutral theory [1] are indicative of some sort of averaging process; we now have much more insight into the likely nature of such averaging (Appendix B). The effect of temporal fluctuations at all levels seems particularly promising as a promoter of idiosyncracy. Niches – in the most general sense – are likely to be flexible, time dependent and with a strong biotic component (see also [9]).

The quasi universal form of observed species abundance distributions should not inhibit looking for the detail seething beneath a rather bland canopy. It might be asked: Given something like a log normal species abundance distribution, can correlations between pairs of species exist? Theoretically, the answer is most certainly yes - Sugihara's [25] sequentially broken stick model contains exactly such correlations. Much more convincing is the observational evidence to be found in Figs 2-4 of [6] (see also [26]): closely related congeners have populations much closer to equality than random pairs drawn from the forest species pools.

This work, building on the work by Pueyo et al [4], has established a general framework for species abundance distributions which covers the full range of ecological models from neutrality to complete idiosyncracy. As such, it successfully achieves the rapprochement between niche and neutrality that has recently been deemed desirable [27].

After the original version of this was posted, S. Pueyo drew our attention to appendix B of [5]. In that paper he applied statistical mechanics to the problem of species abundance distributions and obtained a dynamical origin for a $1/n$ prior by treating the evolution of species populations as diffusive. He pointed out to us that *per capita* birth and death rates are in fact implicit in his appendix B. This had escaped our notice; we are happy to now acknowledge it.

**Appendix A. Some statistical mechanics from atomic physics**

This appendix is designed to give examples of the ideas we have transplanted to ecology in a much simpler physical system; the statistical mechanics of an ideal gas.
This has the advantage of being a very mechanical system, governed by very simple principles which have been known for a very long time and also a system where the applicability is established experimentally far beyond even unreasonable doubt. It is unlikely to be accessible to the general ecological reader.

*A1. The role of a degeneracy factor, with some animadversions on the subject of priors.*

The simplest problem in the statistical mechanics of an ideal gas is to determine the distribution in energy of a very large number of atoms confined within a box such that the energy shared among the atoms is fixed (or the thermodynamic temperature is fixed). In quantum mechanics the energy levels available to an atom confined to a box are quantised and can be visualised as those configurations with standing waves in three orthogonal directions. Thus there are different ways of achieving an energy $E_N$; in fact, the number of ways is equal to the number of ways of choosing three positive integers $n_x, n_y, n_z$ such that the sum of their squares is equal to the integer $N^2$. This is fundamental physics outside of statistical arguments: Jaynes would call it *prior* [1, 2].

If we take these quantum states one by one, not caring whether or not there are other states which share the same energy, then the distribution of a total of $N$ atoms such that there are $n_i$ in level $i$ is determined by maximising the weight factor

$$W = \frac{N!}{n_1! n_2! .... n_k!} \tag{A1}$$

subject to the constraint that the total energy does not change as atoms move around. The result is very simple. The number of atoms occupying a level labelled by $i$ is given by

$$n_i \propto \exp(-E_i / kT) \tag{A2}$$

where $E_i$ is the energy of the particular level labelled by *i*. This is the famous Boltzmann factor (see e.g. [3]). The assumption is that in the absence of the constraint on total energy any atom has the same probability of being found in any of the energy levels (which are discrete, bounded from below but for a gas in a box not bounded from above). Suppose we ask a different question: how many atoms have a particular energy $E$, given that we know that there are $g(E)$ distinct quantum states, all equally probable, with energy $E$?

The answer is the obvious one. Each of the distinct states has occupancy given by (A2) so that the total number of atoms with a given energy $E$ is simply

$$n(E) \propto g(E) \exp(-E / kT) \tag{A3}$$

This is inserting the prior information *a posteriori*.

This result can also be obtained by modifying the weight factor expressed in (A1), inserting the prior information *a priori,* writing instead

$$W^g = N! \Pi \frac{g(E)^{n(E)}}{n(E)!} \tag{A4}$$

where $\Pi$ denotes the continued product over the (quantised) energy levels. When the question is posed in this latter way, we need a *prior* – indeed $g(E)$ contains information outside of maximising the Shannon entropy. We see that not only does the prior depend on the (physical) problem; it also depends on the way the problem is posed!

A question of importance in developing the kinetic theory of gases from statistical mechanics in this way is to ask for the probability density $P(E)$ for an atom to have energy $E$. When taking into account the spacing of the energy levels and degeneracy, the distribution turns out to be given by

$$P(E)dE \propto \sqrt{E} \exp(-E/kT)dE \tag{A5}$$

This yields the velocity distribution of atoms in an ideal gas and has been tested with very great precision. Note that this result might have been obtained by assigning a prior of form $\sqrt{E}$ - even though we are now treating $E$ as continuous. [It might be tempting to adopt the information theory Jeffreys prior, which would be $1/E$. This is however appropriate to minimal prior information and we need the information $g(E)$ from quantum mechanics to describe correctly the ideal gas. This is an awful warning of the dangers inherent in transplanting ideas from statistical inference, perfectly valid in their own domain, into physical (or biological) model building.]

Obviously we could treat the degeneracy problem as assigning to a group of $g(E)$ levels of energy $E$ a different description: they are equivalent to a single level of energy $E$ with a relative probability $g(E)$. Thus if we really had a single level with a dynamical reason for a probability enhanced by $g(E)$ the description would be the same. In the problem of species abundance distributions there is indeed a dynamical reason for a prior probability of $1/n$, as discussed in the text and in Appendix B.

*A2. On the dynamics of an atom, with some remarks on birth and death.*

In contrast to the static picture presented above, this section is intended to illuminate the very complex dynamics which underlie the applicability of statistical mechanics to even a very simple system like an ideal classical gas. We start by writing down an equation governing the population of a given energy level $i$ in terms of the fundamental processes controlling this dynamic situation: atoms enter and leave a level $i$ as a result of collisions with other atoms. The dynamics is governed by the physics of atomic collisions and fortunately the essential points are very simple. Consider a box of gas which contains at some moment a density $N_i$ of atoms in each level labelled $i$. Atoms scatter one from another with a frequency which depends on their density and on their effective size (taken as very small). An atom in state $i$ can be removed by collision with another atom in a state $j$ and similarly an atom can be added to the population of state $i$ as a result of collision between atoms in states $k$ and $l$. Thus we have the equation

$$\frac{dN_i}{dt} = -N_i \sum_{jkl} N_j T_{ijkl} + \sum_{jkl} N_k N_l T_{klij} \tag{A6}$$

where the collisions have of course to be permitted by the relevant conservation laws, most notably conservation of energy (which is why the total energy of all atoms in an isolated box remains constant). In a classical ideal gas the rate factor $T_{ijkl}$ is equal to the rate for the reverse reaction $T_{klij}$ and as a result (A6) is solved for equilibrium by the simple relationship

$$N_i N_j = N_k N_l \qquad \ln N_i + \ln N_j = \ln N_k + \ln N_l \qquad \text{(A7a)}$$
$$E_i + E_j = E_k + E_l \qquad \text{(A7b)}$$

where (A7b) expresses conservation of energy in the scattering process. The two equations (A7a) and (A7b) must be consistent and consequently $N_i$ must be exponentially distributed in the energy $E_i$ (see [3]).[ In (A6) each distinct quantum state has been counted separately but it is easy to modify the equation in terms of $N(E)$ with the inclusion of degeneracy factors.] The first term on the right hand side of (A6) could be whimsically called the death rate for $N_i$ and the *per capita* death rate is the complicated expression $\sum_{jkl} N_j T_{ijkl}$. The second term could be called the birth rate. The relation of the dynamical equation (A6) to an analogue of the master equation (6) in the main text will be explored shortly; first we note that the principles applied above can also be used to treat the imaginary case where the forward and back transition rates are not identical; that is, the physics introduces an energy dependent bias separate from the effects of the constraint of conservation of energy. Suppose that we imagine a case where $T_{ijkl} = E_i^\alpha E_j^\alpha t_{ijkl}$ and where $t_{ijkl} = t_{klij}$. Then the quantity $E_i^\alpha N_i$ is exponentially distributed with $E_i$ and the bias resulting from the *per capita* death rate increasing with $E_i$ is equivalent to a prior $1/E_i^\alpha$. The point is that the nature of the dynamics in a problem of statistical mechanics determines the nature of the prior to be employed.

The problem of species abundance distributions can in fact be expressed in terms of equations directly analogous to (A6, A7). Should a particular species increase in abundance by one individual, then another must decrease in abundance by one individual, for conservation of the total number of individuals. If there are $s_i$ species with abundance $n_i$ then the rate at which species leave the classes $i$ and $j$ is proportional to the product $n_i s_i n_j s_j$, because interactions are going on at the individual level rather than the species level. The analogues of (A7a,b) for species abundance are then

$$n_i s_i n_j s_j = n_k s_k n_l s_l \qquad \ln(n_i s_i) + \ln(n_j s_j) = \ln(n_k s_k) + \ln(n_l s_l) \qquad \text{(A7c)}$$
$$n_i + n_j = n_k + n_l \qquad \text{(A7d)}$$

and hence for these ecological dynamics the quantity $n_i s_i$, rather than $s_i$, is exponentially distributed with $n_i$. In MacArthur's broken stick, it is assumed that the *prior* is a distribution uniform in the number of species $s_i$; the prior for species abundance is uniform in the number of individuals summed over all species in that class, $n_i s_i$.

Finally, we have (A6) look like equation (6) in the text. The easiest way is to write the time dependence of $N_i$ in pieces and focus on the contribution of interactions involving atoms in state *l*. The partial time dependence is then given by

$$\frac{\partial N_i}{\partial t} = -N_i \sum_{jk} N_j T_{ijkl} + N_l \sum_{jk} N_k T_{klij} \qquad \text{(A8a)}$$

which we write as

$$\frac{\partial N_i}{\partial t} = -N_i d_i^l + N_l b_l^i \qquad \text{(A8b)}$$

The complexity of the partial *per capita* rates is evident; the full equation (A6) is a sum over these already complicated equations. If we apply equations (A8) in the equilibrium limit, the ratio $b_l^i / d_i^l$ is, using conservation of energy, simply given by the difference in energy of the states *i* and *l* which we selected

$$\frac{b_l^i}{d_i^l} = \exp\{-(E_l - E_i)/kT\} \qquad \text{(A9)}$$

The ratio is not in general unity and depends exponentially on the difference of the defining quantity energy between the two states considered.

**Appendix B. Dynamical equations for birth and death.**

This appendix is concerned with the structure underlying the averaged equation (6) in the body of the text and in particular the way in which this averaging expresses the idea of idiosyncracy presented by Pueyo et al [1]. We start from the master equation of Volkov et al [2], giving mathematical form to their statement that "The dynamics of the population of a given species is governed by generalised birth and death events…" The equation below is essentially equation (1) of Volkov et al [2] for the time evolution of probabilities $p_{n,k}(t)$ that species $k$ has at time $t$ a population of $n$ individuals:

$$\frac{dp_{n,k}(t)}{dt} = b_{n-1,k}(n-1)p_{n-1,k} - d_{n,k}np_{n,k} - b_{n,k}np_{n,k} + d_{n+1,k}(n+1)p_{n+1,k} \qquad (B1)$$

This differs from (1) of Volkov et al [2] only in that we have defined the $b_n$ and $d_n$ as *per capita* birth and death rates. They may still be functions of $n$ but the dominant factor has been made explicit. It is to be understood that the probabilities on the right hand of (1) are in general functions of time. If (B1) is solved for the steady state or (quasi) equilibrium condition, then each species $k$ is represented by a log series distribution provided that the *per capita* birth and death rates are not dependent on $n$. Each species can have a different log series distribution. If the *per capita* birth and death rates contain some residual $n$ dependence, expressing perhaps the density effects in Volkov et al [3], then the distribution is not quite log series and again each species can have a different distribution. All species have the same distribution in a symmetric model (which in this context includes neutral models). In this case it is immediately obvious that addition of the $S$ equations (B1) for the $S$ species yields equation (6) of the text because the averaging to yield rates $\overline{b_n}, \overline{d_n}$ is trivial.

More interesting is the case where species can have different *per capita* birth and death rates. This does not just imply different numbers; the degree of dependence on $n$ could be different or even the functional dependence. If the differences are slight, obviously the distribution $s_n$ will be close to the symmetric case, but the expression of idiosyncracy in this dynamical context means that the averaging is not trivial.

Again, we start by adding the $S$ equations (B1). The left hand side of the equation resulting from this sum is again $ds_n/dt$ and it is the right hand side which requires particular attention. It is sufficient to focus on (any) one term in the sum; take as an example

$$\sum_k -b_{n,k}np_{n,k}(t) \qquad (B2)$$

where the sum is taken over $k$, all species. If $b_{n,k}$ is independent of $k$ then this term sums to $-nb_ns_n(t)$ and equation (6) for the symmetric case results. More generally, write

$$b_{n,k} = \overline{b_n} + \delta_{n,k} \quad \text{and} \quad p_{n,k} = \frac{s_n}{S} + \varepsilon_{n,k}$$

Substitution of these expressions into (B2) immediately yields

$$-n\{\overline{b_n}s_n + \overline{b_n}\sum_k \varepsilon_{n,k} + \frac{s_n}{S}\sum_k \delta_{n,k} + \sum_k \delta_{n,k}\varepsilon_{n,k}\} \qquad (B3)$$

The first term in (B3) is the corresponding term involving $s_n$ in equation (6) of the full text. The three remaining sums will contain both positive and negative terms; the first two are zero and the last may well be comparatively small. In a full realisation of the notion of idiosyncracy introduced by Pueyo et al [1] "the bits of information [$\delta_{n,k}, \varepsilon_{n,k}$] that are different in different models [$k$] cancel out [in the sums]"; the insertions in square brackets are ours. This is the way that the idiosyncratic limit leads to equation (6) of the text.

So far it has been assumed, at least implicitly, that the birth and death rate parameters are not functions of time. The origins of these parameters are very complicated and they are likely to depend on environmental conditions and on any quasi cyclic variation with time of the probabilities $p_{n,k}(t)$; indeed, the two may not be properly distinguished. In a quasi equilibrium solution to the $S$ equations summed to describe the $s_n$, the terms such as the example in (B2) should be averaged over time on some appropriate scale. Here there are further possibilities for "bits of information to cancel out" and hence we see temporal dynamics as potentially an important promoter of idiosyncracy and hence a major factor in community structure on a scale perhaps not previously envisaged (see, e.g., [4]).